\documentclass[linenumbers]{iopjournal}

\usepackage{amsfonts,amsmath,mathtools,amssymb}
\usepackage[bb=boondox]{mathalfa}

\usepackage{graphicx}
\usepackage{tikz}

\usepackage[symbol]{footmisc}

\usepackage{natbib}

\usepackage{ragged2e}

\newcommand\bb[1]{\mbox{\boldmath{$#1$}}}
\newcommand\grad{\bb{\nabla}}
\newcommand\bcdot{\,\bb{\cdot}\,}

\newcommand\btimes{\,\bb{\times}\,}
\newcommand\rmd{\mathrm{d}}
\newcommand\ompi{\omega_{\mathrm{p},i}}
\newcommand\ompe{\omega_{\mathrm{p},e}}

\newcommand\vecx{\bb{x}}
\newcommand\vecu{\bb{u}}

\newcommand{\vecB}{\bb{B}}

\newcommand{\vecmcS}{\boldsymbol{\mathcal{S}}}
\newcommand{\mcS}{\mathcal{S}}
\newcommand{\mcA}{\mathcal{A}}

\newcommand\be{\begin{equation}}
\newcommand\ee{\end{equation}}

\begin{document}

\articletype{Letter} 

\title{Turbulent Magnetogenesis and Large-scale Magnetic Dynamo Amplification in Ion--electron Plasmas}

\author{Fabio Bacchini$^{1,2,}$\footnote[2]{F.\ Bacchini and F.\ Pucci contributed equally to this work.}\orcid{0000-0002-7526-8154}, Francesco Pucci$^{3,\dagger}$\orcid{0000-0002-5272-5404}, Sergio Servidio$^{4}$\orcid{0000-0001-8184-2151}, Francesco Valentini$^{4}$\orcid{0000-0002-1296-1971}, William H. Matthaeus$^5$\orcid{0000-0001-7224-6024}}

\affil{$^1$Centre for mathematical Plasma Astrophysics, Department of Mathematics, KU Leuven, Leuven, Belgium}

\affil{$^2$Royal Belgian Institute for Space Aeronomy, Solar-Terrestrial Centre of Excellence, Brussels, Belgium}

\affil{$^3$Institute for Plasma Science and Technology, National Research Council (CNR-ISTP), Bari, Italy}

\affil{$^4$Dipartimento di Fisica, Universit\`a della Calabria, Arcavacata di Rende, Italy}

\affil{$^5$Department of Physics and Astronomy, University of Delaware, Newark, USA}

\email{\href{mailto:fabio.bacchini@kuleuven.be}{fabio.bacchini@kuleuven.be}, \href{mailto:francesco.pucci@istp.cnr.it}{francesco.pucci@istp.cnr.it}}


\justifying

\begin{abstract}
\justifying
Using fully kinetic simulations that capture unprecedentedly large (from electron to ion)  scales, we study magnetogenesis driven by continuous large-scale forcing until nonlinear dynamo saturation. We uncover a two-stage mechanism in collisionless ion–-electron plasmas whose dynamics diverge dramatically from the pair-plasma case. In the first phase, electron pressure anisotropy triggers electron-Weibel modes, seeding small-scale magnetic fields. Then, a second growth phase emerges when the more massive ions develop their own strong anisotropy and drive ion-Weibel-type modes; concurrently, a Biermann-battery mechanism contributes to amplifying the magnetic field. This combined dynamics provides a tenfold amplification of the magnetic field in comparison to the pair-plasma case. Over long times, dynamo action continues until the system reaches a statistical steady state. This self-consistent kinetic mechanism provides a plausible explanation for robust magnetogenesis wherever an external forcing continuously stirs the plasma.
\end{abstract}

\section{Introduction}
The problem of magnetogenesis, i.e.\ the creation of magnetic fields, in astrophysical plasmas represents a long-standing open question. Particularly for cosmological-scale environments such as the intracluster medium (ICM), it is expected that ``seed'' magnetic fields of unspecified origin can be dynamo-amplified up to energy equipartition with the local dilute (near-collisionless) plasma medium. However, this dynamics has not been conclusively demonstrated in theoretical models. Fluid- and hybrid-type simulations (\citealt{rincon2016,stongekunz2018,beattie2025,hanebring2026}) and experiments (e.g.\ \citealt{tzeferacos2018}) have provided evidence for large-scale field amplification, but fully kinetic models are required to capture the complete sequence of events (from zero to saturated fields) self-consistently (e.g.\ \citealt{helander2016}).

Mechanisms to generate seeds fields are intrinsically encoded in the Vlasov--Maxwell equations governing collisionless plasmas. Recent works have explored magnetic-field creation in initially unmagnetized, collisionless plasma via Weibel-type (pressure-anisotropy-driven) modes (\citealt{pucci2021,zhou2022,zhou2024,liu2025,liu2026}). \cite{sironi2023} demonstrated the creation and amplification of seed fields to near-equipartition via turbulent dynamo, focusing on the special case of pair plasmas. However, the corresponding ion--electron plasma case, following the system from initial seed creation to dynamo saturation, has not been explored. This case is relevant due to additional effects that only arise when a mass difference between particle species is taken into account (e.g.\ ion-Weibel and Biermann-battery modes), potentially contributing to magnetic-field generation (\citealt{ruyer2015,laishram2024,laishram2026,laishramyoon2026}). 

In this Letter, we explore, for the first time, magnetic-field generation and amplification in an initially unmagnetized ion--electron plasma up to large-scale nonlinear dynamo saturation. Our three-dimensional simulations reach unprecedentedly large spatiotemporal scales, demonstrating that mass separation between plasma species plays a pivotal role in providing further magnetic-field amplification with respect to the pair-plasma case.

\section{Methods and Numerical Setup}
\label{sec:method}

We perform fully kinetic simulations of magnetogenesis with the Particle-in-Cell (PIC) code \textsc{Zeltron} (\citealt{cerutti2013,bacchini2022mri,bacchini2026}). In all cases we start from an unmagnetized, Maxwellian plasma with uniform density $n_0$ and temperature $k_\mathrm{B}T_0$ in a cubic, fully periodic simulation box of side $L$.

To allow for the spontaneous creation of seed electromagnetic fields in the initially unmagnetized plasma, we employ a forcing directly on the particles, representing an external ``kick'' influencing the particle motion. We drive six forcing modes on the macroscopic scale of the simulation box, $\bb{k}L/(2\pi) = (\pm 1,0,0), (0,\pm 1,0), (0,0,\pm 1)$. The forcing is purely solenoidal (i.e.\ no compressive modes are introduced) and is implemented by adding a time-dependent acceleration term to the right-hand side of the particle momentum equation. The Fourier modes associated with this forcing evolve in time according to an Ornstein--Uhlenbeck (OU) process with decorrelation rate $\gamma_\mathrm{corr}\simeq1.5/t_L$, where the turnover time $t_L\equiv L/(2\pi u_\mathrm{rms})$ is determined by the root-mean-square particle speed at steady state $u_\mathrm{rms}$. The acceleration obtained in this way is such that $u_\mathrm{rms}\sim0.2c$. This strategy is akin to that presented by \cite{sironi2023}, and more details on the numerical implementation are discussed there.

Our reference simulation models an ion--electron plasma with a reduced ion-to-electron mass ratio $m_i/m_e=25$ for affordability\footnote{We have also run test simulations up to $m_i/m_e=100$ and found results qualitatively comparable to those shown here.}. The computational domain has linear size $L=500 c/\ompe=100 c/\ompi$, where $\omega_{\mathrm{p},s}\equiv \sqrt{4\pi q_s^2 n_{0}/m_s}$ is the plasma frequency for a particle of species $s$ with charge $q_s$ and mass $m_s$. We employ $768^3$ cells such that the grid spacing $\Delta x\simeq0.65c/\ompe$ resolves the electron skin depth, and 64 particles per cell per species. Ion and electron velocities are initialized from Maxwellian distributions with equal temperature $k_\mathrm{B}T_0$ such that the normalized temperature $\theta_{i,0} = k_\mathrm{B}T_0/(m_i c^2)=(m_e/m_i)\theta_{e,0} = 0.04$. The simulation is run until $t_\mathrm{f}=115 t_L\simeq45{,}700\ompe^{-1}$. For comparison, we have also run a pair-plasma simulation (i.e.\ $m_i/m_e=1$) with exactly the same parameters in terms of system size, initial electron (and positron) temperature, density, and driving modes.

As in \cite{sironi2023}, we limit the maximum particle momentum to $p_\mathrm{max}/(mc)\le0.7$, by rescaling each particle's velocity accordingly when this limit is exceeded while preserving the velocity-vector direction. This is to avoid secular, unbounded heating that results from the continuous OU driving. As noted by \cite{sironi2023}, this strategy implies that at steady state the ratio of turbulent kinetic to internal energy is $\sim0.3$, i.e.\ mildly subsonic turbulence is achieved and in principle maintained indefinitely.

\section{Results}
\label{sec:results}

\subsection{Energetics and Global Evolution}

\begin{figure}
    \centering
    \includegraphics[width=1\linewidth]{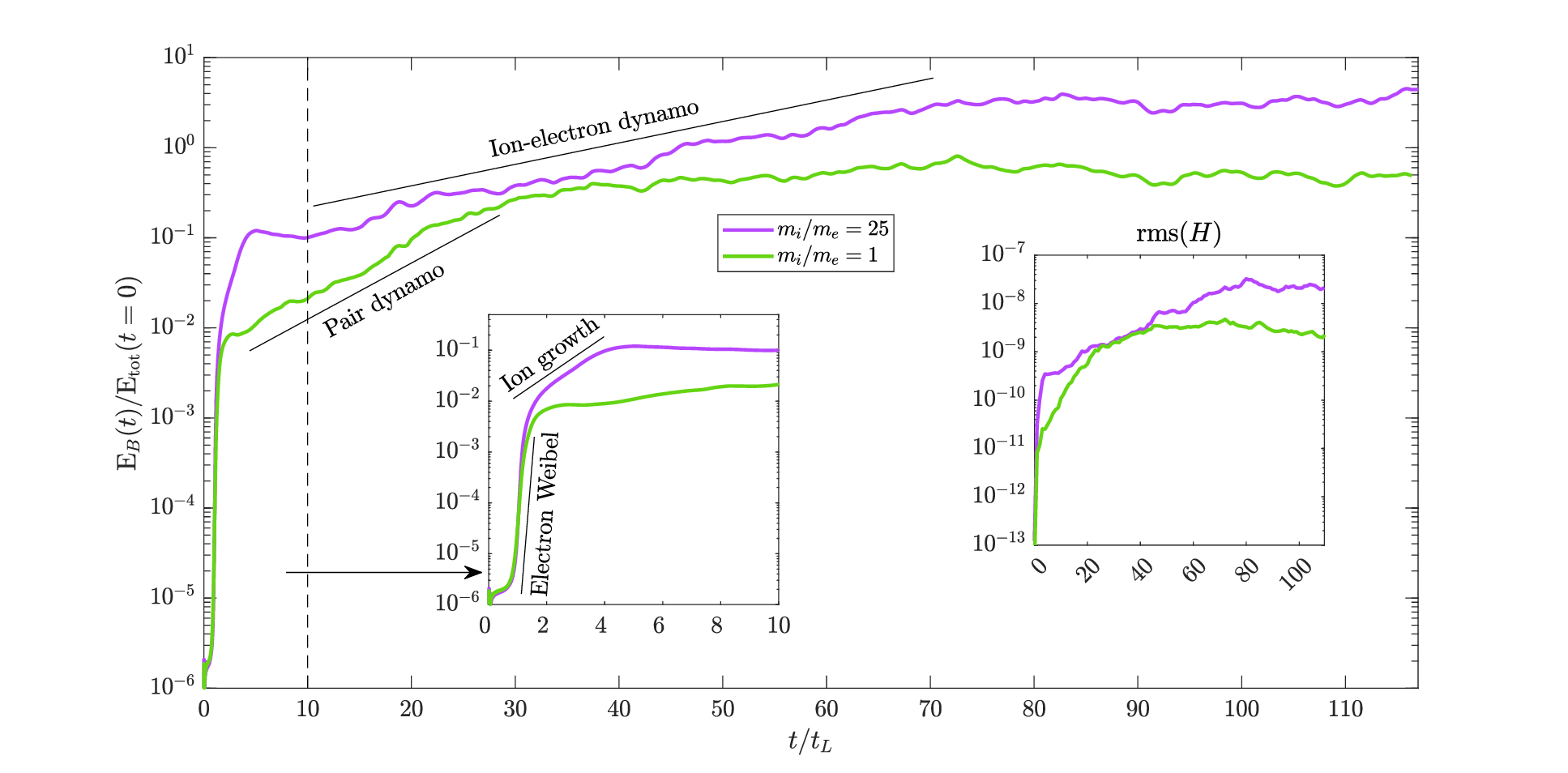}
    \caption{Evolution of the volume-average magnetic energy for the ion--electron (purple lines) and pair-plasma (green lines) runs. In the ion--electron case, the evolution can be divided into an electron-Weibel phase ($t<1.5t_L$), an ion-growth phase ($1.5t_L<t<4.5t_L$), an ion--electron dynamo phase ($4.5t_L<t<75t_L$), and a quasi-steady phase ($t>75t_L$). The early-time phases are zoomed into in the left inset. The right inset shows the evolution in time of the root-mean-square magnetic helicity for the two runs.}
    \label{fig:enB_H}
\end{figure}
\begin{figure}
    \centering
    \includegraphics[width=1\linewidth,trim={11mm 19mm 0 0},clip]{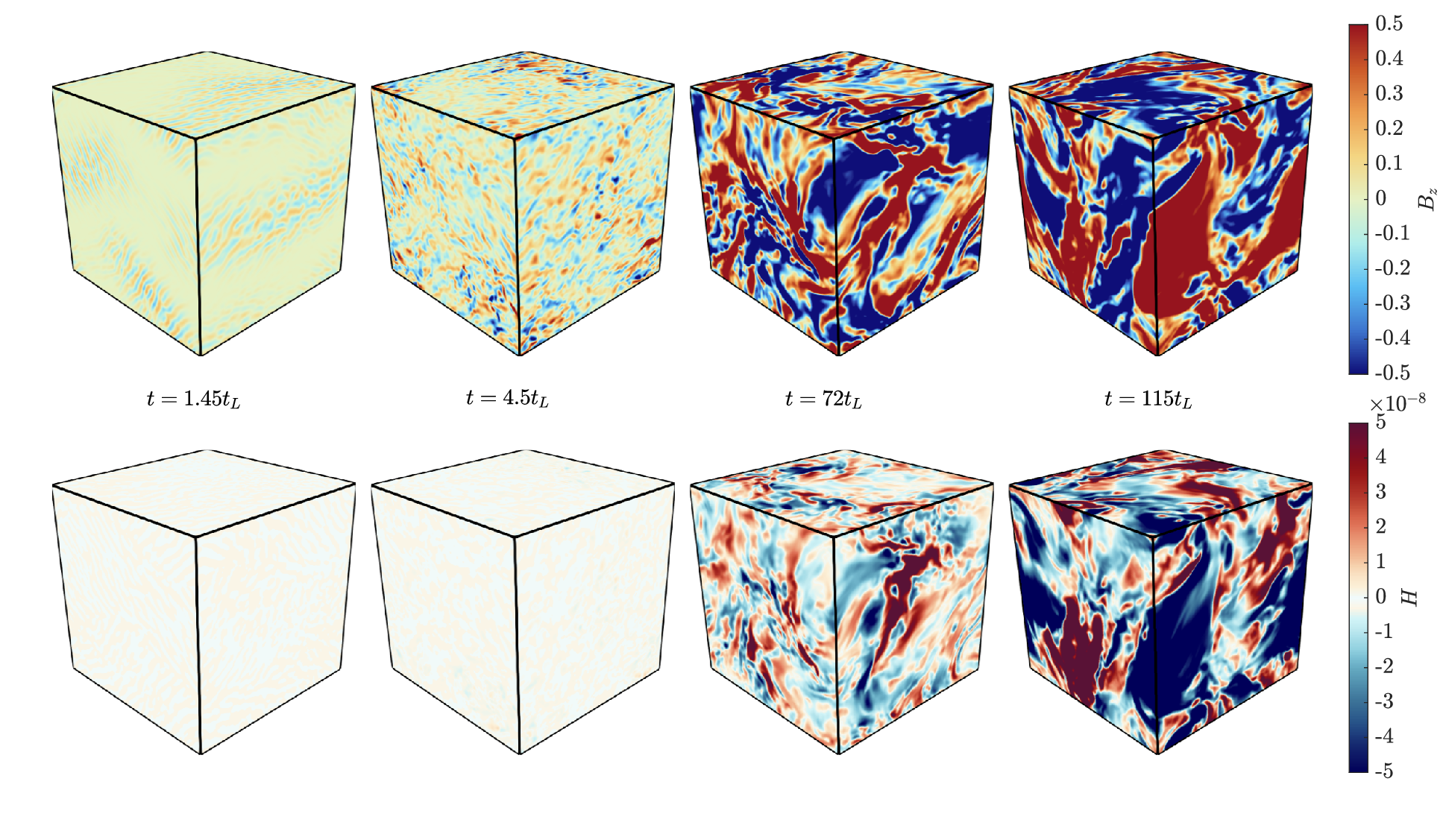}
    \caption{Representative snapshots of the spatial distribution of the magnetic field ($z$-component, top row) and magnetic helicity (bottom row) during the four evolution phases of the ion--electron run.}
    \label{fig:fields_H_horiz}
\end{figure}

The global evolution of the system is shown in Fig.~\ref{fig:enB_H}, where we show the volume-integrated magnetic energy over time, normalized to the initial total energy, for our reference run ($m_i/m_e=25$) compared with the pair-plasma run ($m_i/m_e=1$). For the reference run, the evolution can be divided into four main phases (the first two phases can be better appreciated in Fig.~\ref{fig:enB_H}, left inset):
\begin{itemize}
    \item \textit{Electron-Weibel phase:} As soon as the simulation starts, the OU forcing causes counteraligned streams of electrons (which are lighter); this is a sufficient condition to immediately drive electron-Weibel modes, growing rapidly, producing magnetic fields, and saturating at $t\simeq1.5t_L$.
    \item \textit{Ion-growth phase:} Ions react to the forcing slower than electrons, but are similarly driven toward counterstreaming motions. However, in contrast to the electron dynamics of the previous phase, ion-driven modes develop on top of a state where magnetic fields are nonzero, and where electron modes are also subdominant but still active. We label this an ``ion-growth'' phase, where magnetic energy grows by a further factor of 10 albeit at a rate $\sim10$ times slower than during the initial electron-Weibel phase, saturating at $t\simeq4.5t_L$.
    \item \textit{Ion--electron dynamo phase:} With the seed fields created by electron and ion dynamics, a dynamo action ensues and slowly drives a magnetic-energy increase by another factor $\sim50$, saturating at $t\simeq75t_L$.
    \item \textit{Steady-state phase:} After $t\simeq75 t_L$, the system has achieved equipartition between kinetic and magnetic energy (see next section), and magnetic-energy growth is halted while a steady state is reached and maintained indefinitely.
\end{itemize}
In contrast, in the pair-plasma run, we can identify only three phases\footnote{\cite{sironi2023} distinguish four phases also for the pair-plasma case, by differentiating between Weibel onset and filament-merging phases. Here, we do not make this distinction.} (electron Weibel, dynamo, and steady state). The most striking feature of the ion--electron evolution is the ion-growth phase ($1.5t_L\le t\le 4.5t_L$), which is completely absent in the pair-plasma case. Through this and subsequent phases, at the end of the run the ion--electron simulation achieves 10 times more magnetic energy than the pair-plasma case. Note that the energy injected is not the same in the two runs: our charge- and mass-independent forcing causes particles of different mass to experience the same acceleration, implying that more massive particles will receive higher amounts of energy.

In Fig.~\ref{fig:fields_H_horiz} (top row) we show reference snapshots of the spatial distribution of the magnetic-field component $B_z$ at representative moments during each phase. From left to right, we observe: i) The emergence of small-scale magnetic fluctuations during the electron-Weibel phase ($t= 1.45t_L$); ii) Larger- (ion-)scale fluctuations appearing during the ion-growth phase ($t=4.5t_L$); iii) The creation and merging of system-size magnetic filaments during the dynamo phase ($t=72t_L$); and iv) System-size, persistent structures of coherent $B_z$ during the steady-state phase ($t=115t_L$). Over time, we indeed notice the appearance of larger and larger coherent-field regions, which suggests an increase in magnetic helicity (an ideal nondissipative invariant in both the MHD and Hall-MHD frameworks; see e.g.\ \citealt{servidio2008}). To quantify this dynamics, typical of turbulent-dynamo evolution (e.g.\ \citealt{pouquet2019}), we compute the magnetic helicity $H=\bb{A}\bcdot\bb{B}$, where $\bb{A}$ is the vector potential ($\bb{B}=\grad\btimes\bb{A}$), and display its root-mean-square value over time (Fig.~\ref{fig:enB_H}, right inset) and its spatial distribution at the same times shown for $B_z$ (Fig.~\ref{fig:fields_H_horiz}, bottom row). Here, we observe a steady increase in average (rms) magnetic helicity as the system evolves through the main dynamic phases, and the presence of regions of high magnetic helicity spatially correlated with regions of coherent $B_z$. This is consistent with MHD expectations, where $H$ typically undergoes an inverse cascade and creates large-scale structures. In comparison with the pair-plasma case, the ion--electron simulation achieves a larger average magnetic helicity, and in particular a sharp increase during the ion-growth phase which does not occur in the $m_i/m_e=1$ case.

\begin{figure}
    \centering
    \includegraphics[width=1\linewidth]{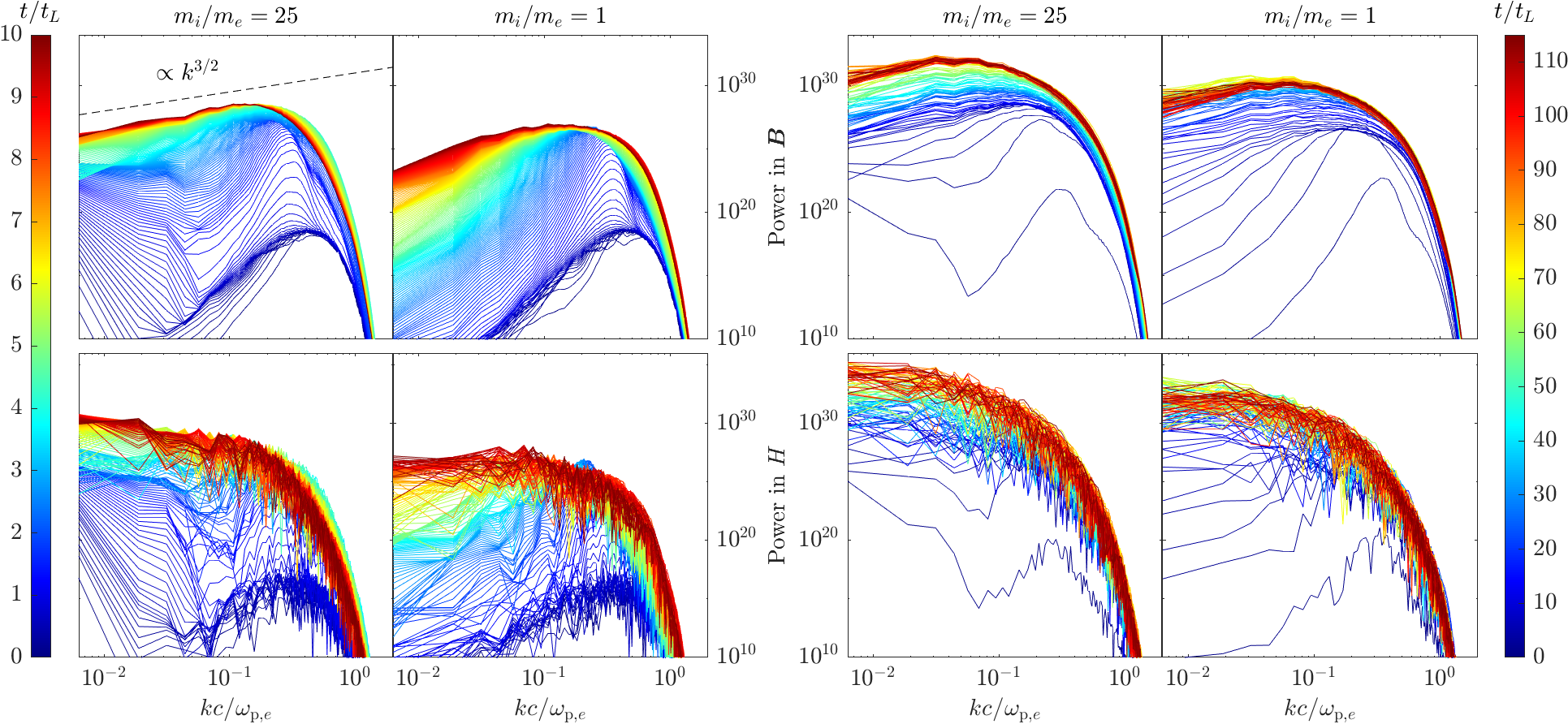}
    \caption{Isotropic power spectra of $\vecB$ (top row) and $H$ (bottom row) during the early-stage evolution (left side; $t\in[0, 10]t_L$, cadence $0.05t_L$) and during the whole run (right side; $t\in[0, 115]t_L$, cadence $1t_L$) for the ion--electron and the pair-plasma cases. The Kazantsev scaling $\propto k^{3/2}$ is shown for reference in the top-left panel.}
    \label{fig:fft}
\end{figure}

To better understand the observed magnetic-energy growth during the different phases, in Fig.~\ref{fig:fft} we show isotropic power spectra of $\bb{B}$ and $H$ over time. The left columns show high-cadence spectra (every $0.05t_L$) between $t=0$ and $t=10t_L$, covering the electron-Weibel and ion-growth phases. In the ion--electron case, we observe that energy in $\vecB$ starts growing around $k\simeq0.3\ompe/c$, consistent with theoretical estimates for electron-Weibel modes (e.g.\ \citealt{schoefflersilva2020}), and progressively shifts to larger scales as the electron-Weibel phase proceeds and ends, as also observed by \cite{pucci2021}. During and after the electron-Weibel phase, we notice power accumulating at scales both above ($k\gtrsim0.6\ompe/c$) and below ($k\simeq0.03\ompe/c$) the electron-Weibel scale. This secondary energy growth is absent in the pair-plasma case, suggesting that it is intrinsically tied to the presence of ions. Finally, we observe that the spectrum progressively evolves toward a Kazantsev scaling $\propto k^{3/2}$ (\citealt{kazantsev1968}) which is reached at the end of this early stage of the evolution.

\begin{figure}
    \centering
    \includegraphics[width=0.55\linewidth]{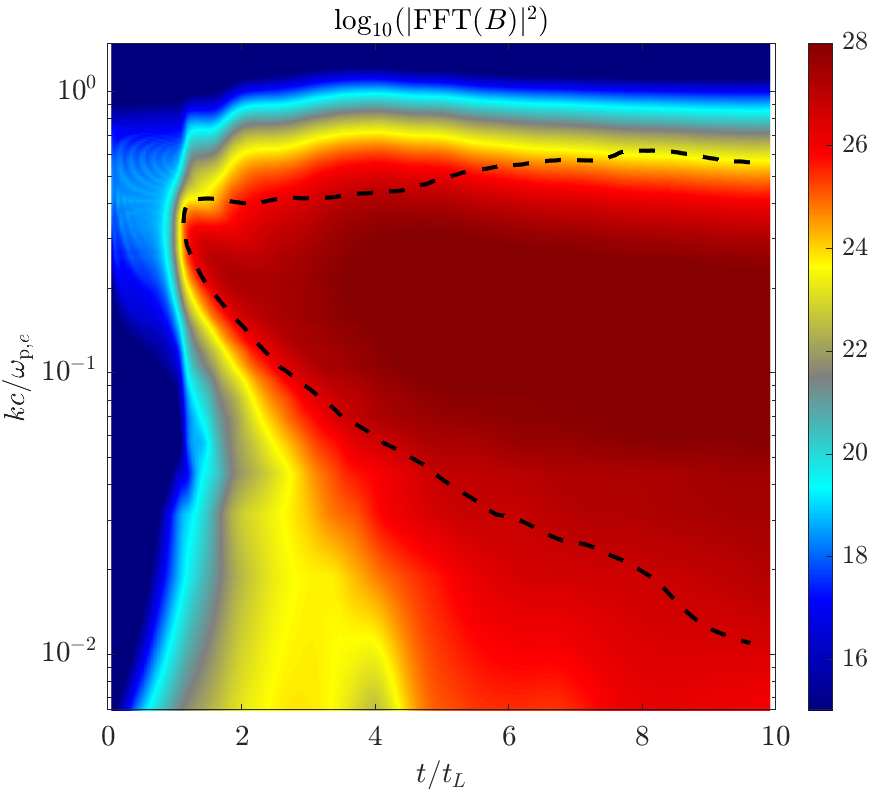}
    \caption{Power in $\bb{B}$ over time for each $k$ during the electron-Weibel and battery phases of the ion--electron run. The dashed black line represents one contour of the corresponding result for the pair-plasma case passing through $(k,t)=(0.3\ompe/c,1.5t_L)$, the electron-Weibel reference wavenumber at the time of electron-Weibel saturation. This highlights additional growth in magnetic energy above and below the electron-Weibel scale during the ion-growth phase of the ion--electron case.}
    \label{fig:fft2D}
\end{figure}

The right columns of Fig.~\ref{fig:fft} show power spectra at low cadence (every $t_L$) between $t=0$ and $t=115t_L$, i.e., until the end of the run. We observe that the ion--electron case accumulates, in total, more power than the pair-plasma case, but the overall spectrum is fairly similar over long times. At the end of the run, in both cases, the magnetic field has developed up to the largest (box-size) scales, due to dynamo action. Finally, in the bottom row of the same figure, we also show magnetic-helicity spectra at the same times as those of $\vecB$. Similar considerations apply here: at early stages, power in $H$ grows at scales above and below the electron-Weibel scale, and over long times we observe slightly more power for the ion--electron case accumulating up to the largest admissible scales. The propensity for magnetic helicity to accumulate at the longest wavelength is well-known in MHD \citep{frisch1975}. The difference between the ion--electron and pair-plasma cases during the early phases can be more clearly observed in Fig.~\ref{fig:fft2D}, where we show the power in each $k$ over time.

From these analyses, it appears that the ion--electron dynamics plays a crucial role in determining both total and scale-wise magnetic-energy growth in comparison with the pair-plasma case. The main candidate explanation for this difference may lie in the fact that ions and electrons are decoupled and behave differently, due to their different masses, throughout the system's evolution. We discuss this in more detail in the next section.

\subsection{Anisotropy, Battery Effects, and Early- vs.\ Late-stage Dynamics}

\begin{figure*}
    \centering
    \includegraphics[width=1\linewidth]{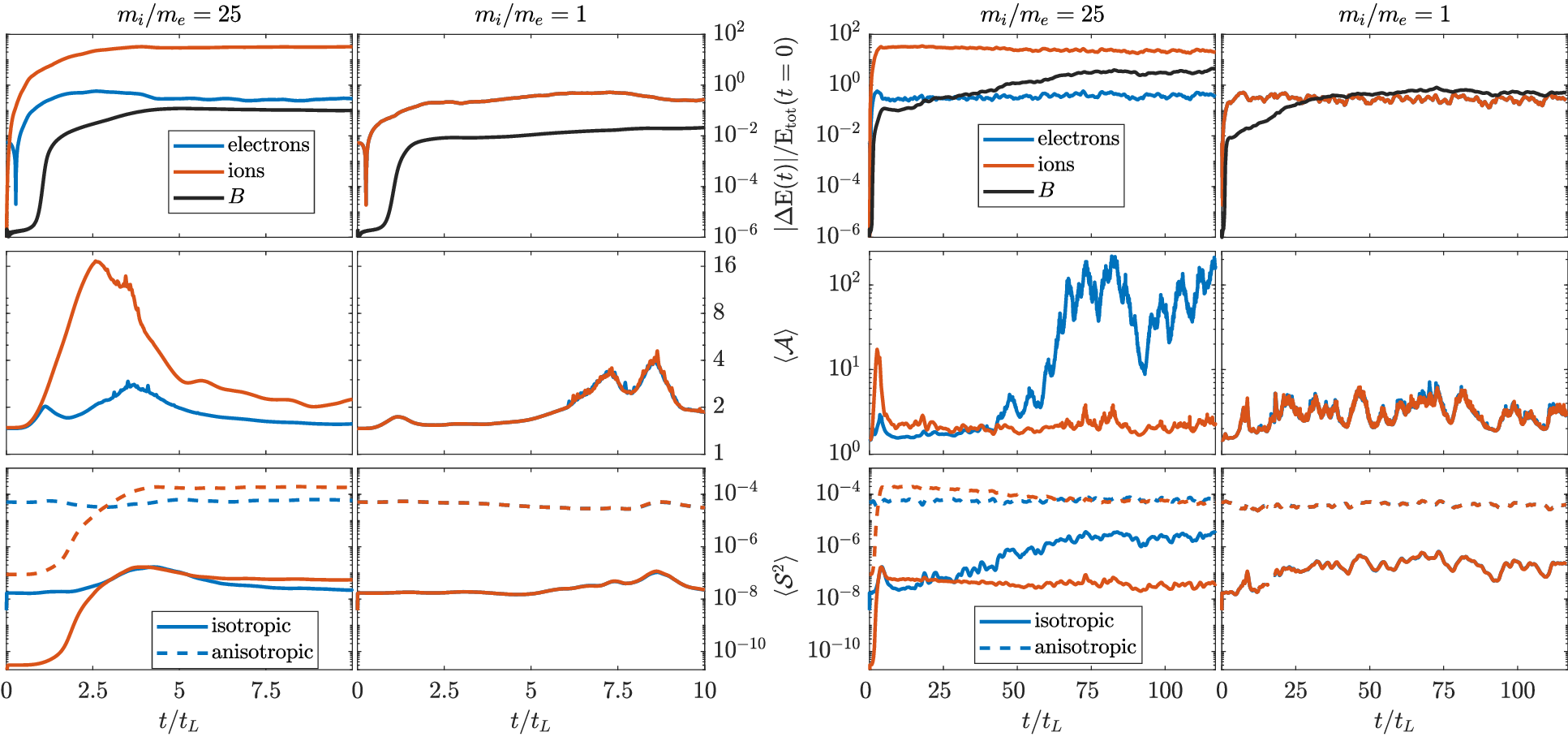}
    \caption{Volume-averaged diagnostics for the early-time ($t<10t_L$; left half) and late-stage (until the end of the run; right half) evolution of the pair-plasma and ion--electron cases. Top row: kinetic and magnetic energy. Middle row: anisotropy as defined in Eq.~\eqref{eq:anisotropy}. Bottom row: battery terms as defined in Eqs.~\eqref{eq:Siso} and \eqref{eq:Sani}}
    \label{fig:int_quant}.
\end{figure*}

To analyze differences in the evolution of ions and electrons, we provide two global measures. 

First, we quantify pressure anisotropy for each species. Here, we consider the approach of \cite{servidio2012} and \cite{pucci2021}. Since there exists no preferential direction for the magnetic field (which is even completely absent at $t=0$) in our runs, we define anisotropy as the local lack of thermal-pressure uniformity along different spatial directions. We measure this by computing, for each particle species $s$, the eigenvalues of the pressure tensor $\textbf{P}_s \equiv m_s\int (\widetilde{\vecu}\widetilde{\vecu}/\widetilde{\gamma}) f_s(\vecx,\widetilde{\vecu},t) \rmd^3 \widetilde{\vecu}$, where $\widetilde{\gamma}\equiv\sqrt{1+\widetilde{u}^2/c^2}$ and $\widetilde{\vecu}$ is the particle velocity in the frame moving with bulk velocity $\bb{V}_s\equiv(1/n_s)\int(\vecu/\gamma) f_s(\vecx,\vecu,t) \rmd^3 \vecu$. At each time step, we calculate the eigenvalues $\lambda_3<\lambda_2<\lambda_1$ of $\textbf{P}$ for each species and obtain a measure of the anisotropy
\be
\mathcal{A}\equiv\frac{2\lambda_1}{\lambda_2+\lambda_3},
\label{eq:anisotropy}
\ee
where $\mcA\simeq1$ corresponds to a completely isotropic situation\footnote{In all of our runs, at $t=0$ we measure $\mcA\gtrsim1$ even though we initialize isotropic velocity distributions, due to inevitable numerical noise in PIC simulations.}. Note that, while $\mcA$ can capture the early phases of anisotropy generation (when no magnetic field is present), the actual anisotropy with respect to the local magnetic field or the local strain-tensor eigensystem represents matter for future analysis.

In addition, we measure the presence of magnetic-energy sources from canonical-battery effects. In a collisionless plasma, the canonical vorticity related to each plasma species $s$, $\bb{Q}_s = m_s \grad\btimes\bb{V}_s - q_s\bb{B}$ (e.g., \citealt{laishram2024}), evolves according to
\be
\frac{\partial\bb{Q}_s}{\partial t} = \grad\btimes(\bb{V}_s\btimes\bb{Q}_s) \underbrace{- \grad\btimes\left(\frac{\grad\bcdot\textbf{P}_s}{n_s}\right)}_{\vecmcS_s}.
\label{eq:battery}
\ee
As explained by \cite{laishram2024}, the first term on the right-hand side represents the frozen-in condition for $\bb{Q}$ and the second term ($\vecmcS$) is a genuine source term of canonical vorticity, and therefore of $\bb{B}$, from plasma thermodynamics. It is clear that, in an initially unmagnetized, immobile plasma, the only source of $\bb{Q}$ (therefore $\bb{B}$) must come from pressure/density inhomogeneities. In particular, we can distinguish between a purely isotropic contribution,
\be
\vecmcS_{s,\mathrm{iso}} = - \grad\btimes\left(\frac{\grad p_s}{n_s}\right),
\label{eq:Siso}
\ee
where $p\equiv\mathrm{tr}(\textbf{P})/3$, and a purely anisotropic contribution
\be
\vecmcS_{s,\mathrm{ani}} = - \grad\btimes\left(\frac{\grad\bcdot(\textbf{P}_s-p_s\textbf{I})}{n_s}\right).
\label{eq:Sani}
\ee
The isotropic term is usually associated with Biermann-battery effects, and the anisotropic term with Weibel-type effects. We measure these contributions throughout the system's evolution as a proxy for magnetic-field generation by thermodynamic effects.

The results are displayed in Fig.~\ref{fig:int_quant}. We show the evolution in time of the volume-averaged kinetic and magnetic energy (top row), pressure anisotropy $\mcA$ (middle row), and magnitude of $\vecmcS_\mathrm{iso}$ and $\vecmcS_\mathrm{ani}$ (bottom row). We distinguish the early evolution phases (i.e.\ the initial $10t_L$, including electron-Weibel and ion-growth phases; left half of the figure) from the overall evolution during the whole run (right half of the figure), and show both the pair-plasma and the ion--electron cases. Unsurprisingly, in a pair plasma the two species behave practically identically, while in the $m_i/m_e>1$ case, ion and electron dynamics differ in all measured quantities. The electron response is generally faster, but ion-related quantities typically achieve larger values.

We first focus on the early-time behavior ($t<10t_L$, left half of Fig.~\ref{fig:int_quant}). Concerning anisotropy, we notice that electrons achieve a first maximum during the time period identified as the electron-Weibel stage. This occurs both in the pair-plasma and in the ion--electron cases, suggesting that a similar dynamics takes place for the lighter particles. Anisotropy in the electron pressure tensor is indeed the prime candidate to fuel the electron-Weibel growth in $\vecB$, and the visible correlation between $\mcA_e$ and magnetic-energy growth supports this interpretation. Ions, instead, peak at a (much) larger $\mcA_i$ precisely during the ion-growth phase, indicating that their slower dynamics is, during this stage, the main driver of $\vecB$ growth. It is also interesting that, during the ion-growth phase, electrons appear to follow the ion dynamics and mildly increase their $\mcA_e$, likely responding to ion-driven modes. Anisotropy in both species rapidly decreases after the ion-driven growth in $\vecB$ saturates.

The early-time evolution of $\vecmcS$ reveals several interesting features. First, the anisotropic contribution is always dominant, both in the pair-plasma and in the ion--electron cases. This is expected since our solenoidal driving force mostly creates velocity shears, therefore preferentially driving phase-space anisotropy. In the ion--electron case, the electron battery terms dominate the initial electron-Weibel phase; during early-time evolution, however, the ion contribution increases up to the point of equaling, and eventually surpassing, the electron one, both for isotropic and anisotropic terms. The magnitude of $\vecmcS_i$ becomes dominant and saturates specifically around $t=4.5t_L$, i.e.\ precisely during the ion-growth phase, again suggesting that ion modes dominate magnetic-field creation. While anisotropic terms remain dominant, in the ion--electron case the isotropic contribution is up to 10 times larger, during the ion-growth phase, than in the pair-plasma case. This aspect is discussed in more detail in the next section.

The overall and late-time behavior is shown in Fig.~\ref{fig:int_quant} (right half). While the pair-plasma simulation saturates relatively quickly and displays no striking differences (in any quantity) between early and late times, the ion--electron run shows very active dynamics during the dynamo phase and subsequent evolution. After the electron and ion anisotropy is quenched at the end of the ion-growth phase, we can observe a long period (until $t\simeq50t_L$) of relative isotropy in both species. At $t\simeq75t_L$, a very rapid increase in $\mcA_e$ occurs in correspondence with the end of the dynamo phase. The electron anisotropy here achieves and maintains very large values up to $\mcA_e\sim100$, while the ion anisotropy remains essentially stable at low values. In this phase, magnetic energy is no longer growing appreciably (although the forcing is still active), and the system has achieved a steady state. Therefore, the measured electron anisotropy here does not appear to cause Weibel-driven net magnetic-field generation.

The time evolution of the battery terms over long timescales shows that ion contributions, initially dominant, progressively lose importance. While $\mcS_\mathrm{ani}$ remains much larger than $\mcS_\mathrm{iso}$ for both species, at the end of the run $\mcS_{i,\mathrm{ani}}\lesssim\mcS_{e,\mathrm{ani}}$ and $\mcS_{i,\mathrm{iso}}\sim\mcS_{e,\mathrm{iso}}$, with the latter having increased by two orders of magnitude throughout the dynamo phase. Electron dynamics therefore appears to be very active during the late-stage evolution of the system, and this must be due to the electron response to the global state driven by ions. Indeed, the equivalent pair-plasma case displays no such late-time behavior, indicating that the difference in mass between species is playing a key role in the dynamo and steady-state evolution of the ion--electron case.

These results highlight several important points. The most striking features of the ion--electron run are the qualitatively different behavior of the two species in terms of anisotropy and contribution to magnetic-field generation. In comparison with a pair-plasma case in which the two species behave essentially identically, here we observe an additional magnetic-energy growth phase and a very active electron dynamics during the late stages of the evolution.

\section{Discussion and Conclusion}

\begin{figure*}
    \centering
    \includegraphics[width=1\linewidth,trim={0mm 12mm 0 0},clip]{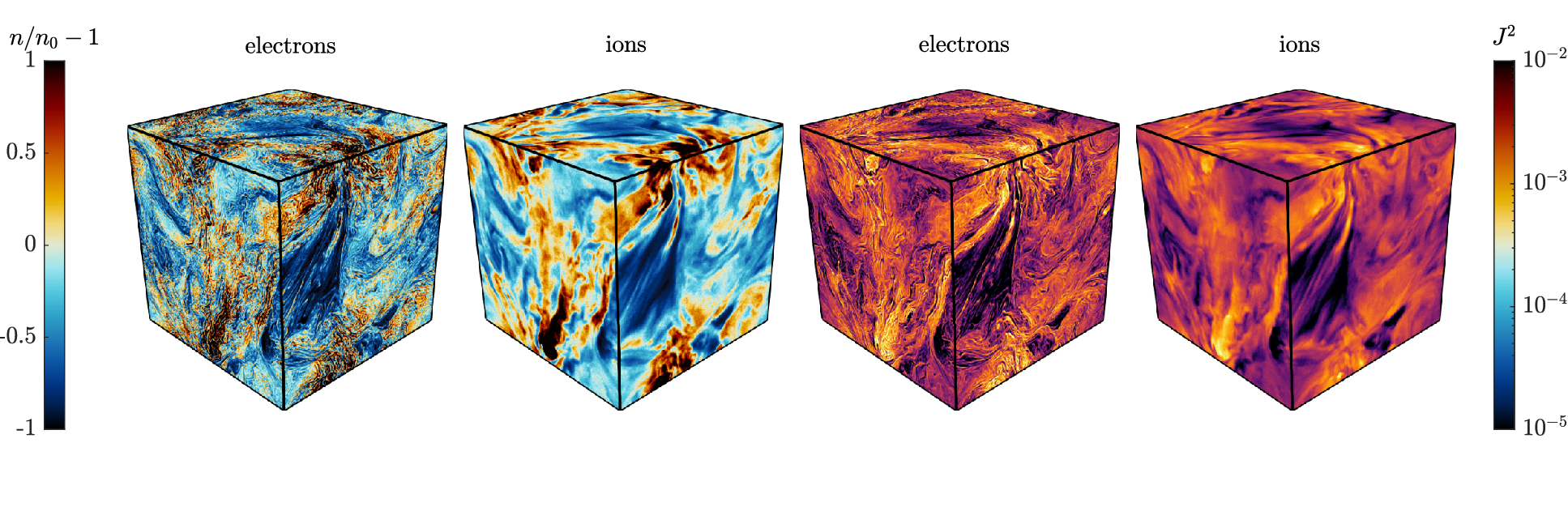}
    \caption{Spatial distribution of number density (left) and current-density magnitude (right) for electrons and ions at $t=115t_L$.}
    \label{fig:dens_J}
\end{figure*}

We have conducted three-dimensional fully kinetic simulations of magnetogenesis in an initially unmagnetized ion--electron plasma, covering an unprecedented range of spatial and temporal scales. We observe the evolution of the system through several characteristic phases, from initial Weibel-driven magnetic growth to statistical steady state. In comparison with pair-plasma simulations (previously explored also by \citealt{sironi2023}), we observe dynamics that can only be attributed to a finite ion-to-electron mass ratio and that result in several previously unreported effects.

The early-time system evolution ($t<10t_L$) displays an electron-Weibel initial phase, observed also in the pair-plasma case, where electron dynamics dominates particularly in terms of anisotropy-induced modes (which our forcing strategy preferentially drives). This first phase is followed by an additional magnetic-growth phase that is absent in pair-plasma runs. This phase is characterized by:
\begin{itemize}
    \item A decoupling in the ion and electron pressure anisotropy, with ions reaching a much higher average $\mcA_i$;
    \item A progressive increase in ion-battery activity, with both isotropic- and anisotropic-battery effects eventually dominated by ions;
    \item A subsequent (but weaker) follow-up electron reaction to the ions, with electrons mildly increasing their $\mcA_e$ and isotropic-battery effects in response to the ion's dominating dynamics;
    \item A further (but slower) 10-fold increase in magnetic energy over the value reached at the end of the electron-Weibel stage.
\end{itemize}
Based on our diagnostics, we can attribute this early-time evolution to two potentially complementary mechanisms.

The increase in ion anisotropy during this phase indicates that the more massive ions may, at this point, have been subjected to enough forcing to create large-scale counterpropagating flows. These can result in ion-Weibel modes, slower than electron-Weibel ones, which further amplify magnetic fluctuations. The growth in spectral power at large $k\simeq 0.6\ompe/c$ (i.e.\ at scales larger than the electron-Weibel scale) visible in the magnetic-energy spectra (Figs.~\ref{fig:fft} and \ref{fig:fft2D}) may in fact support this hypothesis\footnote{The growth in spectral power at small $k\simeq0.03\ompe/c$, as well at the system-size scale, instead, is consistent with electron-slippage dynamics recently found in fully kinetic simulations (\citealt{pusztai2026}).}. Indeed, because ions achieve very large anisotropy, the scale of ion-Weibel growth decreases drastically and may even shrink below that of electron modes. Note, however, that this ion-Weibel mode cannot develop in the same way as the earlier electron-Weibel one (which started from a state of zero magnetic field), because at the moment when ion modes emerge the underlying magnetic-field configuration has been heavily modified by the electron dynamics. This shrinking in ion-Weibel scales due to anisotropy has indeed been described by \cite{ruyer2015} precisely in the case of ion-mode growth in the presence of an electron-modified magnetic background, and is compatible with what we observe here.

At the same time, we measure a net increase in battery effects during this early phase, which can contribute to magnetic-energy evolution in our ion--electron case. In a pair plasma, the sum of $\vecmcS_{s}$ among all species is in principle exactly zero; only very small differences between electrons and positrons survive, with anisotropic-battery terms amplifying these differences more easily via Weibel-type instabilities. In our ion--electron case, where all particles are driven with the same mass-independent acceleration, we observe that the two species decouple almost immediately. Pressure and density gradients do not have to be equal for ions and electrons, and any difference between the species' thermodynamics will reflect in a net contribution to battery effects. In Fig.~\ref{fig:dens_J} (left) we show the spatial distribution of electron and ion density at $t=115t_L$, during the steady state; we observe appreciable density differences between the two species, as well as strong, ubiquitous density gradients. These develop precisely during the ion-growth phase and become larger over time, and can therefore directly contribute to magnetic-energy growth via isotropic and anisotropic battery effects. In particular, Fig.~\ref{fig:int_quant} (bottom row) shows that the absolute maximum in the average $\mcS_{i,\mathrm{iso}}$ is reached during the ion-growth phase. This implies that, on top of ion-anisotropy effects (included in $\mcS_{i,\mathrm{ani}}$), Biermann-battery effects are also present and contribute to magnetic-field generation. Finally, note that the moment when $\mcS_{i,\mathrm{iso}}\sim\mcS_{e,\mathrm{iso}}$ ($t\simeq4.5t_L$) corresponds precisely to the end of the ion-growth phase. Since the two $\vecmcS_s$ contribute with opposite signs to creating $\vecB$ (see Eq.~\eqref{eq:battery}), this is indeed the moment when net Biermann effects are quenched and this growth phase ends.

The late-time behavior, from $t\simeq50t_L$ onward, also displays extreme differences between pair-plasma and ion--electron cases. At this time, the pair-plasma run has already achieved equipartition and entered a steady state, while the ion--electron simulation is still evolving through a long dynamo phase. While ions remain quiescent, anisotropy here starts to increase sharply, achieving extreme ($\mcA_e>100$) values; interestingly, this is not associated with an increase in $\mcS_{e,\mathrm{ani}}$, implying that this anisotropy does not contribute to (anisotropic) battery effects. The isotropic $\mcS_{e,\mathrm{iso}}$, conversely, increases by up to 3 orders of magnitude during the dynamo phase, also in stark contrast with the pair-plasma dynamics. In summary, in this phase we observe active Biermann-battery effects (contributing to magnetic-energy growth) strongly linked with electrons, which may contribute to the larger total magnetic energy achieved in the ion--electron case with respect to the pair-plasma run. We also observe the development of strong electron anisotropy \emph{not} contributing to anisotropic-battery effects, which deserves further explanation.

In Fig.~\ref{fig:dens_J} (right), we show the spatial distribution of the electron and ion current-density magnitude at $t=115t_L$, during the final steady state. We observe that, while both species display regions of strong currents, the electron dynamics is much more active and decoupled from the ions, with myriad small-scale current sheets filling the whole volume and a general presence of more numerous current filament structures. This suggests that active, primarily electron-driven, turbulent magnetic reconnection is occurring. Reconnection typically tends to drive pressure anisotropy (e.g.\ \citealt{comisso2024}), which is compatible with the measured increase in $\mcA_e$ at late times. Therefore, we conjecture that turbulent reconnection is dominating the electron physics during late-time system evolution, driving anisotropy which is however not linked with (anisotropic) battery effects. Reconnection dissipates magnetic fields generated by dynamo action, eventually reaching an effective balance between gain and loss of magnetic energy. We have checked that, at this stage, the average ion Larmor radius is comparable with the box size, which supports the idea that the ion dynamics is effectively unimportant and electrons are the main active species during the late-stage evolution.

In summary, our results show, for the first time, that i) ion--electron physics generally fuels stronger magnetic-field dynamo via and additional growth phase mediated by ion and battery effects, and ii) an ion--electron system with an external forcing achieves a late-time steady state where electron-mediated magnetic reconnection balances magnetic-field creation. These results substantially differ from the previously explored pair-plasma dynamics and have important consequences for theoretical models of ion--electron plasmas subjected to external forces. For example, in the ICM, gravitational disturbances can act as an effective forcing driving magnetic-field generation \citep{hitomi2016}; in astrophysical accretion disks, the disk's differential rotation naturally introduces a continuous shear acting as a forcing onto the plasma, which may contribute to dynamo mechanisms such as the nonlinear  magnetorotational instability (\citealt{balbushawley1991,riquelme2012,bacchini2022mri,bacchini2024mri,sandoval2024,gorbunov2025apjl}). In the future, we will carefully explore larger parameter spaces to definitely establish the role of, e.g., the separation between different species and between system size and kinetic scales, the precise hierarchy of modes contributing to magnetic-energy creation during the different phases evolution, the different possible natures of the applied forcing, etc. This work, however, already demonstrates that ion--electron dynamics plays a fundamental role in creating magnetic fields in various astrophysical scenarios, and ought to be taken into account in theoretical models.

\ack{F.B.\ would like to thank Muni Zhou, Istv\'an Pusztai, Lorenzo Sironi, and Martin Lemoine for useful discussions during the development of this work.}

\funding{
F.B.\ acknowledges support from the FED-tWIN programme (profile Prf-2020-004, project ``ENERGY'') issued by BELSPO. F.B.\ and F.P.\ acknowledge support from the FWO Junior Research Project G020224N granted by the Research Foundation -- Flanders (FWO). 
This research  partially supported by the U.S.\ National Science Foundation, award PHY-210883.
The computational resources and services used in this work were provided by the VSC (Flemish Supercomputer Center), funded by the Research Foundation - Flanders (FWO) and the Flemish Government.
}


\data{Data presented in this work can be shared upon reasonable request.}




\end{document}